# IMPROVING ENERGY EFFICIENCY IN WIRELESS SENSOR NETWORKS THROUGH SCHEDULING AND ROUTING

RATHNA. R[1] AND SIVASUBRAMANIAN. A[2]

[1]Research Scholar, Sathyabama University, Jeppiaar Nagar, Chennai-119, TamilNadu, India

`rathna08@yahoo.co.in`

[2]Professor and Head, Department of Electronics and communication, St.Joseph's College of Engineering, Jeppiaar Nagar, Chennai-119, TamilNadu, India

`shiva_31@yahoo.com`

## ABSTRACT

*This paper is about the wireless sensor network in environmental monitoring applications. A Wireless Sensor Network consists of many sensor nodes and a base station. The number and type of sensor nodes and the design protocols for any wireless sensor network is application specific. The sensor data in this application may be light intensity, temperature, pressure, humidity and their variations .Clustering and routing are the two areas which are given more attention in this paper.*

## KEYWORDS

## 1. INTRODUCTION

As the wireless Sensor Networks are made up of tiny energy hungry sensor nodes, it is a challenging process to retain the energy level of those nodes for a long period. They are equipped with limited computing and radio communication capabilities. This work is on the attempt to reduce the power consumption of nodes, by concentrating on the radio, which has four states of operations at various time intervals.

Wireless Sensor Networks are made up of tiny sensors which are used for monitoring or sensing data. Because of their small size, power supply is provided by a small battery, which, when deployed in a 'not-easily reachable' place, cannot be replaced or recharged frequently. The purpose of all these nodes is to monitor the required data and send them to a Base station which may be in a remote place.

These Wireless Sensor Networks are deployed in so many areas like vital signal monitoring in tele-homecare systems, ecology monitoring which are widely used for monitoring wild-life, rare-micro organisms, changes in the sea or lake water, soil after natural disasters like typhoon, tsunami, flood and soil erosion, monitoring climatic changes, structural monitoring for e.g. Monitoring the conditions of a bridge after its construction, monitoring the historic buildings and Surveillance in Defence organizations.

Storage mechanism is also very simple and can only provide limited space. So acquisition of precise data and immediate transfer of the data to sink node is very important. Data processing and data transfer require more power. When, the data has to be transferred and when, it needs to be stored depends on the state of the radio in the node. To conserve energy, we can switch the radio to sleep state when there is no data to send or receive. This method of making the radio to be in sleep state and making it active if any event is detected is called as on-demand scheme or event-based scheme.

There is another method of scheduling i.e. on regular time interval all the nodes will be either in sleep mode or active mode. This is synchronous scheme. But the overhead of maintaining all the nodes in the synchronized state becomes complex. It is not necessary to keep all the nodes active at one time. WSN can follow a scheduling pattern, accordingly, at any instant; we can make only a limited number of nodes active. The work presented in this paper is based on this type of asynchronous mechanism.

## 2. SCHEDULING

As a sensor node in WSN is small, its power supply unit should be very small and also it should support all its operations without degrading the performance. The communication protocol used should be light weight and it should not consume more energy. Hence, we are going for a good scheduling protocol and while applying it, power consumption is the one which should be kept in mind.

Any Scheduling protocol will keep only a subset of nodes to be in active state and keeping others in-active or in sleep state. A scheduling protocol will be the best if it keeps only a minimum number of nodes active at any instant. There are so many scheduling protocols available. For a WSN, a scheduling protocol should use narrow band modulation techniques. Low data transfer rate is enough for a WSN in Environmental monitoring applications.

The TDMA (Time Division Multiple Access) based scheduling protocols make the nodes to be in inactive mode, until their allocated time slots. The TDMA based protocols [1] are designed such that the shortest path for communication will be found out and only a particular link will be in wake up mode for a transmission.

Any scheduling protocol for the WSN for Medium Access Control (MAC) should have the following:

- Narrowband Modulation techniques.
- Good throughput efficiency.
- Moderately low Data transfer rate.
- Less Hardware complexity.
- Low access delay, low transmission delay and low overhead.

The TDMA based scheduling allocates separate time slot for each node to access the medium to send the sensed data or to forward the aggregated data.

Some of the already existing scheduling protocols are:

- The B-MAC [12] is a MAC protocol which introduces the first Low power Listening (LPL) Protocols. But it does not support all types of Radio.
- The WiseMAC is an improved one, which reduces the length of the preamble by sending the data only to its neighbors. But this also faces the same problem of B-MAC.
- The SpeckMAC consists of SpeckMAC-B and SpeckMAC-D. It also aims at reducing the energy consumed for sending the data to active nodes by reducing the preamble size.
- The performance of X-MAC is similar to B-MAC, with the additional fixed length preamble using advertisement cycles for sending the data.
- The D-MAC protocol staggers or sends the 'send and receive' time slots for single packet exchange, to all the available paths to its neighbors.
- MIX-MAC protocol makes the WSN to adapt to different types of MAC protocols depending on the situations like packet size.
- Inverse Log Scheduling- centralized and distributed protocols[14] assigns long transmission times for those sensor nodes which face worse channel conditions.
- The centralized and de-centralized sensor scheduling protocols as given in [8] concentrate on both power conservation and coverage. They are designed for military surveillance purpose.
- Sift [12] is a MAC protocol which is designed on the basis of sending the important or high priority information first, with less delay and then sending the low priority information.

### 3.1. Sleep/wake-up scheduling

For a asynchronous type of sleep/wake-up scheduling periodically only certain nodes have to wake-up and send or receive. All the other nodes should be in sleep state.

While going for a switch from active to sleep state, and then from sleep to active state, the condition to be checked is that the energy it consumes for the switch should be low when compared with the energy it consumes when it is always in the active state.

$$E_{\text{wasted in switching}} < E_{\text{Saved}}$$

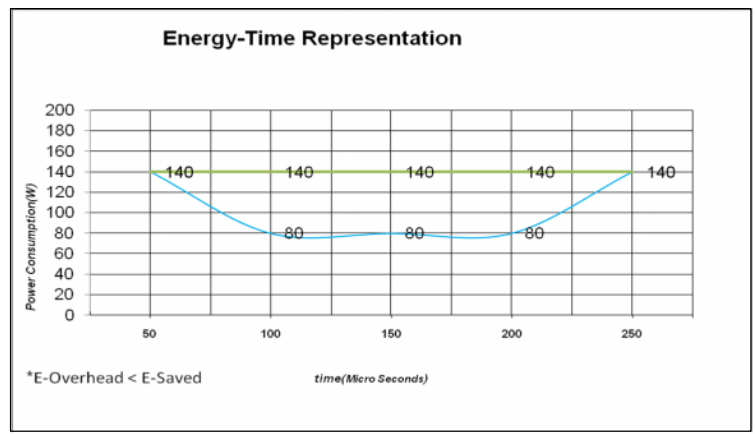

Figure 1. Graph depicting change in power consumption by using a sleep/wake-up scheduling

This scheduling protocol is designed based on the radio of the node.

- $r_s$ -Radio in sleep state
- $r_t$ -Radio in transmitting state
- $r_r$ -Radio in receiving state
- $r_l$ -Radio in listening state

The energy conumed for the switch can be represented by $E_{sl}$(sleep to listen), $E_{st}$(sleep to transmit) and $E_{sr}$(sleep to receive).

For a time slot, say 't', if the node is in sending state, it will be denoted by $r_{t,t}=1$. If it is not in sending state, it will be denoted by $r_{t,t}=0$. So at a particular instant say t, a node can be in any one of the four states. So we can state the condition like this

$$r_{t,t}+ r_{r,t}+ r_{l,t}+ r_{s,t} =1$$

For the state switch from sleep state to listening state, the energy consumed can be calculated by $t (r_{s,t}+ r_{l,t}+1 ) E_{sl}$.

During every cycle, the radio goes to all the four states. When the radio is either sending or receiving or simply listening, it is said to be active. If it goes to sleep, it is said to be in sleep state.

The Energy wasted in simply switching between states $E_{wasted}$ in switching can be calculated by

$$E_{\text{wasted in switching}} = t_{s\text{-}a} (p_{active} + p_{sleep})/ 2$$

Table 1. Notations used

| Notations | Descriptions |
|---|---|
| $r_s$ | Radio in sleep state |
| $r_t$ | Radio in transmitting state |
| $r_r$ | Radio in receiving state |
| $r_l$ | Radio in listening state |
| $E_{sl}$ | Energy consumption during the switch from sleep state to listen state |
| $E_{st}$ | Energy consumption during the switch from sleep state to transmitting state |
| $E_{sr}$ | Energy consumption during the switch from sleep state to receiving state |
| $p_{active}$ | Power consumed in active state($r_t, r_r, r_l$) |
| $p_{sleep}$ | Power consumed in sleep state $r_s$. |
| $t_{s\text{-}a}$ | Time taken for going to active state( $r_t, r_r, r_l$) from sleep state. |
| $t_{s\text{-}a}$ | Time taken for going to sleep state from active state( $r_t, r_r, r_l$). |
| $t_a$ | Time at which the radio becomes active as per the schedule to send or receive any data. |
| $t_s$ | Time at which the radio decides to go to sleep state as per the schedule. |

The Energy saved because of this switching can be calculated by

$$E_{Saved} = (t_a - t_s) \, p_{active} - (t_{a-s}(p_{active} + p_{sleep})/2 + (t_a - t_s - t_{a-s}) \, p_{sleep}$$

In this work, the basic scheduling is designed based on this first check i.e.

$$E_{wasted\ in\ switching} < E_{Saved}$$

Then only switching between active and sleep states will be beneficial. The notations used in the above expressions are described in the Table I.

## 3. CLUSTERING

If the number of nodes used in the application is large, then data aggregation has to be done. If all the nodes try to send the sensed data to the BS(Base Station), more energy will be consumed, eventually more nodes will die frequently. The data gathered by a set of nodes has to aggregated and sent to the BS from that point. A tree like arrangement of wireless sensor nodes is used in this work. All the nearby nodes are grouped to form different clusters.

This idea is inspired from the work of LEACH [13] (Low Energy Adaptive Clustering Hierarchy). In this algorithm, different set of nodes become the cluster heads each time. Every time the node which is the cluster head takes the responsibility of aggregating the data from its nearby nodes and sends the data to the BS, there by reduces the energy wastage of all the nodes.
There are also other types of routing techniques available. Most of them are derived from the LEACH [13]. The other types are Random walk [15] protocol-In this the nodes are simply arranged like a grid. The nodes are assumed to be present at the grid junctions, and then the desired route is found out. There are again three more subdivision under these. But practically for the environmental monitoring applications, this type of keeping the nodes at grid junction i.e., the topology becomes impossible.

In Directed Diffusion method [14], the query will be broadcasted from the node, it will reach only the active (alive) nodes. The interested nodes will then send back the data to the desired node. In turn this will lead to lot of energy wastage, since broadcasting needs lot of energy.

So the cluster based routing is best suitable routing protocol for Environmental monitoring applications. In this work, the clusters are formed based on a weight attached to each node. While forming the clusters, the following rule should be followed- No two clusters should have one or more nodes in interference range. Interference range is that , the two nearest nodes in two different clusters should not be in either transmitting(rt )or receiving state(rr). This will cause interference and overhearing of packets and thereby wastage of energy.

For all the nodes in each cluster, certain weight (w) is added based on the time, it senses or receives the data from other node. All the nodes in each cluster will wake up based on the weight of the cluster. At that moment the nodes will either send or receive data to or from the other nodes respectively. This weight to the cluster is a group of time slots to all the nodes in that cluster.
So in a tree, there will be clusters with different weights. The cluster with greatest weight (smallest time stamp) will be allotted the first available timeslot. Here weight is added based on the information received by the nodes from other nodes and environment. In a tree like structure, the nodes in the clusters have to receive the data from its child nodes and have to send that data packet to the parent node (the node which is in the higher order of the sending node in the tree hierarchy). The time slots will be allocated to the nodes in a decreasing order of weight to the nodes in the tree structure. If this mechanism is followed for scheduling the nodes in a WSN for environmental monitoring applications, surely the overall energy consumption of the WSN can be reduced by a considerable range.

## 4. SIMULATION AND RESULTS

The above described method of Cluster based sleep/wake-up scheduling is tested in a simulated WSN and it proves to be efficient. Network Simulator-2(NS2) is used in this work for simulation.NS2 is one of the best simulation tools available for Wireless sensor Networks. We can easily implement the designed protocols either by using the otcl coding or by writing the C++ Program. In either way , the tool helps to prove our theory analytically.

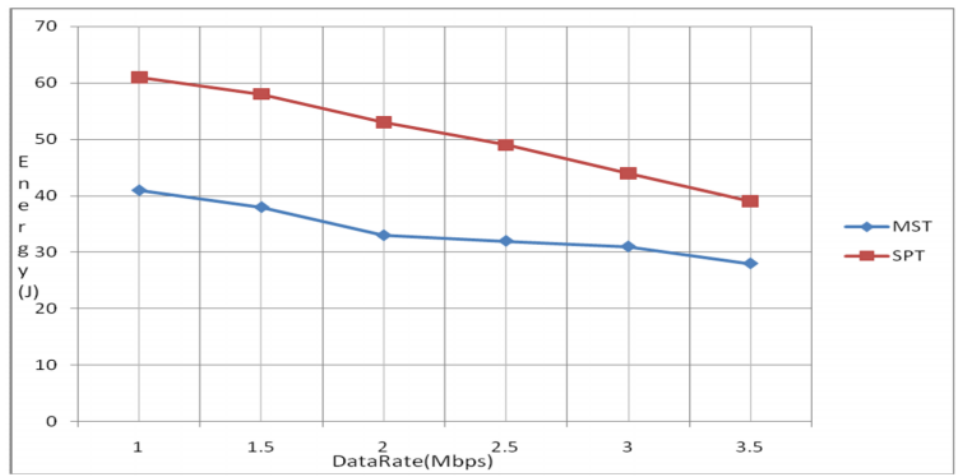

Figure 1. Graph depicting the comparison of Energy consumption by SPT and MST

The WSN is simulated by applying the clustering technique to the tree structures formed by BFS, Shortest Hop path Tree(SPT) and Minimum Spanning Tree(MST). The clustering of nodes in a MST has shown the best result and it is depicted in the figure.

When this scheduling algorithm is implemented it gives a good improvement in the reduction of time delay as well as the overall energy consumption.

## 5. CONCLUSION

The proposed algorithm in this paper is completely TDMA based. It helps to reduce the energy consumption by reducing the number of times; a node has to wake up, during a time slot, to be in active mode. The underlying concept in this paper is efficient usage of energy. It has been proved also. The time delay is also reduced to a small extent. The future work can be done by combining both the TDMA and FDMA based slot allocation.

## Authors

**Rathna.R** has a master in Information Technology from Sathyabama University, Chennai. Working as Assistant Professor in the Department of Information Technology, Sathyabama University, She is pursuing Research in the field of Wireless sensor Networks. Her Research interest lies in Energy optimization in Wireless Sensor Networks.

**Sivasubramanian.A** holds a doctorate in Optical Networks. He is now working as Professor and Head of the Department of Electronics and Communication, St.Joseph's college of Engineering, Chennai. His areas of interest are Optical fibre transmission, Optical Networks, Mobile communication networks, Wireless sensor networks.